# Investigation of III-V nanowires by plan-view transmission electron microscopy: InN case study

Esperanza Luna[1,*], Javier Grandal[1,3], Eva Gallardo[1,2], José Manuel Calleja[2], Miguel Ángel Sánchez-García[3], Enrique Calleja[3] and Achim Trampert[1]

[1]Paul-Drude-Institut für Festkörperelektronik, Hausvogteiplatz 5-7, D-10117 Berlin, Germany

[2]Departamento de Física de Materiales, Universidad Autónoma de Madrid, E-28049 Madrid, Spain

[3]ISOM and Departamento Ingeniería Electrónica, ETSI Telecomunicación, Universidad Politécnica de Madrid, E-28040 Madrid, Spain

*Email: luna@pdi-berlin.de; Tel: + 49 30 20377 281; Fax: +49 30 20377 515

**Abstract.** We discuss on the observation by plan-view high-resolution transmission electron microscopy (TEM) of InN nanowires (NW). The main difficulties arise from the scarce suitable methods available for the plan-view specimen preparation. We explore different approaches and find that the best results are obtained using a refined preparation method based on the conventional procedure for plan-view TEM of thin-films and specifically modified for the NW morphology. The fundamental aspects of such a preparation are the initial mechanical stabilization of the NWs and the minimization of the ion-milling process after dimpling the samples until perforation. The combined analysis by plan-view and cross-sectional transmission electron microscopy of the NWs allows determining the degree of strain relaxation and reveals the formation of an unintentional shell layer (2-3 nm thick) around the InN NWs. The shell layer is found to be composed of bcc-$In_2O_3$ nanocrystals with a preferred orientation with respect to the wurtzite InN: $In_2O_3$ [111] ∥ InN [0001] and $In_2O_3$ <110> ∥ InN <11-20>.

INTRODUCTION

Self-assembled nanowires (NWs) based on III-V semiconductors are being the object of increasing attention for basic physics and novel applications. In order to study the properties of the NWs, it is frequently desirable to determine their three dimensional shapes and interfaces, which requires the combined analysis of the information extracted from cross-sectional and plan-view observations. Representative examples are core-shell NW heterostructures (Tambe et al., 2010; Wang et al., 2012; Dayeh et al., 2013; Rudolph et al., 2013; Zheng et al., 2013; Dimakis et al., 2014; Lenrick et al., 2014) or the analysis of the surface electron accumulation layer at the sidewalls of InN NWs, which has been a subject of intense debate in the last years (Calleja et al., 2007; Lazić et al., 2007; Segura-Ruiz et al., 2009, 2010; Werner et al., 2010; Zhao et al. 2012, 2013).

While plan-view observation by scanning electron microscopy (SEM) is a widely extended practice, there is a comparatively smaller number of works on the plan-view observation of NWs by transmission electron microscopy (TEM). In general, plan-view TEM of NWs has a higher potential in answering questions about shell structures, characteristics of the sidewall facets or the existence of lateral gradients in composition, for instance. The reduced number of investigations on this subject is mainly due to the limited and sophisticated specimen preparation methods for plan-view observations of such NWs. Recently, the use of advanced fabrication techniques based on microtomy (Tambe et al., 2010; Zheng et al., 2013) or focus ion beam (FIB) (Schreiber et al., 2012; Lenrick et al., 2014) have been proposed. Despite the specific advantages of each technique, these are, however, not suitable for all materials neither for all NWs dimensions. The high-energy ions from FIB, for instance, may amorphize the material, aside from producing undesirable amorphous coatings. Also most of the works on TEM specimen preparation by FIB or microtomy refer to NWs with diameters larger than 300 nm. TEM plan-view investigation of NWs with diameters smaller than 100 nm (otherwise a representative dimension for III-V NWs grown by molecular beam epitaxy, MBE) still seems to represent a big challenge. Furthermore, as deduced from the published data, most of the NWs-based plan-view TEM specimens do not fulfill the basic requirements for high-resolution TEM (HRTEM) observations. Very high-quality specimens with minimized damage, controlled thickness (5-50 nm) and suitable orientation (i.e. the NW axis parallel to the electron beam) are thus mandatory.

In this work, we propose an alternative approach to get HRTEM-quality plan-view TEM specimens of NWs which is based on the conventional procedure of thin-film plan-view TEM specimen preparation but specifically modified for the NW morphology. We illustrate the potential of our proposal by showing case studies scientific questions (requiring HRTEM observations and further structural analysis) on the highly-sensitive InN material, grown as NWs on silicon substrates by MBE.

MATERIALS AND METHODS

**Sample growth**

The InN NWs were grown by plasma-assisted (PA) MBE on both Si(111) and Si(001) substrates, either on bare Si substrates or on a high temperature (HT) AlN buffer layer. The MBE system used in this work was equipped with a radio frequency (rf)-plasma source providing active nitrogen and standard Knudsen cells for In and Al. After a standard degreasing, the substrates were heated in the growth chamber at 900°C for 30 min to remove the native oxide. The growth temperature was measured with an optical pyrometer calibrated with the $1 \times 1 \rightarrow 7 \times 7$ surface reconstruction transition temperature of Si(111) at 860 °C (Suzuki et al., 1993). Growth on HT-AlN-buffered Si was preceded by depositing 10 monolayers of metallic Al at 870ºC. Afterwards a 50 nm thick AlN buffer layer was grown at the same temperature to improve the epitaxial alignment (Grandal et al., 2007). Upon the growth of the HT-AlN buffer layer, the growth was stopped to decrease the temperature down to 475°C, the optimal one for InN (Grandal & Sánchez-García, 2005). In the case of the growth of InN NWs on bare Si, we explored the effect of different growth temperatures between 450 and 480 °C. Thicker NWs and even coalesced layers were obtained as the growth temperature was reduced.

**Cross-section TEM specimen preparation and investigation**

The morphology of the as-grown samples was routinely investigated using SEM. The NWs are hexagonal prisms with their axis oriented parallel to the growth direction, c-axis, i.e. [0001]. However, depending on the exact growth conditions, there are some variations on the NWs morphology. The average diameter of the NWs is 80-120 nm with a height ranging from about 400 nm to 1 µm. TEM images were obtained with a JEOL JEM-3010 microscope (JEOL Ltd., Tokyo, Japan) operating at 300 kV and equipped with a GATAN CCD camera.

Cross-sectional TEM specimens of semiconductor NWs are successfully prepared using conventional techniques, where the NWs are glued for mechanical stabilization and thinned by mechanical grinding, dimpling and Ar-ion milling (Trampert et al., 2004). On the contrary, the plan-view TEM preparation of NW specimens presents several difficulties. The main one concerns the mechanical properties associated with the morphology of NW samples. NWs are fragile objects and they easily break or detach from the substrate if an improper manipulation is done. In order to increase their consistency, a previous mechanical stabilization of the NWs would then be a crucial prerequisite, for instance, by filling the space between the NWs with a cured epoxy or resist, as will be discussed later.

Indeed, in general, the most common way to analyze NWs is to take advantage of the easy detachment of the NWs from the substrate for their observation by TEM. In that case, the NWs are removed from the native substrate after sonication in an isopropanol bath or directly after scratching from the sample surface and then dispersed on a TEM carbon grid. This preparation method has the advantage that the NWs are free of the artifacts introduced during the conventional TEM preparation process, where the final Ar-ion milling sometimes results in an undesired amorphous coating on the sample surface. In the investigation of NWs prepared this way, it is clear that the observed features are sample-related and that they cannot be an artifact introduced during the preparation. However, there are serious disadvantages of this method: (i) only very thin NWs (with the appropriate thickness, 5-50 nm) are suitable for investigation and (ii) the geometry of the lying NWs (on the TEM carbon grid) has to be the right one for plan-view observations (i.e., with the electron beam parallel to the wire axis). From the practical point of view, the dispersed NWs on the TEM grid exhibit a random distribution of positioning and orientations, making tedious and complicated their observation in plan-view geometry, since most of the NWs lie on the grid along their c-axis (i.e., with the electron beam perpendicular to the wire axis). Representative examples are presented in figure 1. Although TEM operation allows for specimen tilting, the allowed tilting angle is limited to about ± 20°, thus hindering the proper alignment of the NWs. Nevertheless, because of its simplicity, this is the most widely used technique for the observation of NWs in the TEM. For cross-section TEM observations of NWs, we use either this technique or the standard preparation procedure (glued together, grinding, and ion milling).

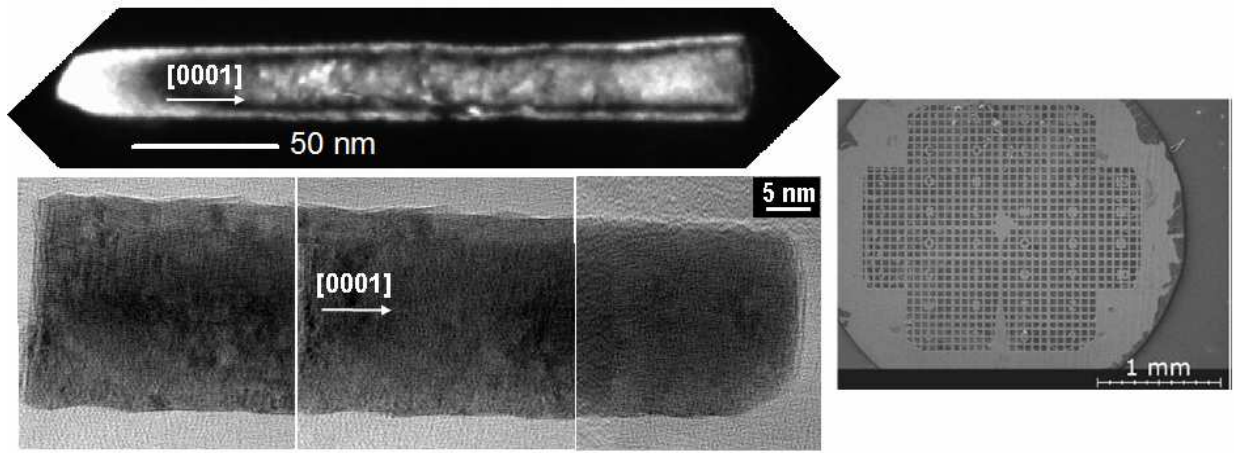

**Figure 1.** Cross-section TEM micrographs displaying representative examples of InN NWs lying on a TEM carbon grid after being removed from the substrate by sonication.

**New approach in the specimen preparation procedure for plan-view TEM**

We explored an innovative preparation procedure of plan-view TEM specimens. Although more elaborate, it is based on the conventional tools and steps for plan-view TEM specimen preparation of thin-films, conveniently modified to deal with the NW morphology. As already mentioned, the first necessary step consists in the mechanical stabilization of the NWs. Although the idea of filling out the inter-wire space is theoretically very simple, in reality there are several inconveniences. The first and fundamental one is related to the method of depositing a thin and uniform coating layer that matches the average height of the NWs (500-1000 nm thick), with material stable against electron irradiation and compatible with the high vacuum operating conditions at the TEM. Our first attempt consisted in coating the samples with different epoxies and photo-resists, namely we employed those used in technological processing (optical and electron lithography, metallization). In addition, to improve the homogeneity of the coated layer, samples were also spun. Several experiments were performed varying the rotation speed of the spinner, as well as the sample temperature, that was heated up during and/or after spinning. Other coating tests without spinning were also investigated. In all the previous cases we got unsatisfactory results, where non-uniform coatings, irregular wetting of the surface or too thick coatings were obtained.

Good results were achieved, however, when using a hand-coating procedure with the cured epoxy GATAN G1, which is stable under electron-beam irradiation. Depending on the size of the sample we deposited small glue droplet(s) on the sample surface [we deposit 4-5 droplets of about 0.3 mm in diameter for a 2 mm x 2 mm sample, cf. figure 2(a)] that spreads over the surface by supervised heating (the sample is kept under observation while heated on a hot-plate at about 80 °C) resulting in a thin (~ 500 nm) coating layer, filling out completely the space between the NWs. The final thickness of the coating layer depends on the size and distribution geometry of the initial droplet(s) and is slightly smaller than the average height of the NWs [figure 2(c)], thus we can check by visual inspection that the coating thickness closely matches the NWs height. Once the NWs were protected with this coating layer, the procedure continued with the conventional mechanical lapping and dimpling [figure 2(d)-(e)]. The second innovative approach was introduced in the dimpling process and consisted in allowing the dimpling continues until sample perforation occurred [figure 2(f)]. Finally, the sample was introduced into the GATAN precision ion polishing system (PIPS) and sputtered with low energy ions (2 keV) impinging from the substrate side for no longer than 10 min [figure 2(g)]. In the case here studied, InN-based samples, the final Ar-ion milling process is critical, since the ion bombardment promotes the removal of In atoms and enhances In-clustering, resulting in severe sample damage and the introduction of structural artifacts. This sputtering time reduction significantly improves the quality of the TEM specimens. The thickness of the TEM specimens prepared using this approach perfectly fulfills the requirements for high quality HRTEM as has been observed in this work and plan-view HRTEM micrographs of NWs with diameters smaller than 100 nm are successfully obtained (cf RESULTS AND DISCUSSION). Furthermore, prepared in this way, we obtain thin, large plan-view sections and hence gain access to a large number of different NWs (> 200) within one single TEM specimen, thus allowing statistical TEM investigations. An additional advantage with respect to previous methods is that here we investigate the originally preserved as-grown samples that have not suffered any alterations due to the detachment process from the substrate. Since the samples keep their original structure, the method also allows for the investigation of the nanowire-substrate interface providing complementary information to that obtained from cross-sectional TEM. Finally, we found that the most reproducible results were obtained if the height of the NWs was close to 500 nm. Longer nanowires (~1 µm) can be also prepared using this method; however, additional difficulties in the reproducibility of the required thickness for HRTEM were found.

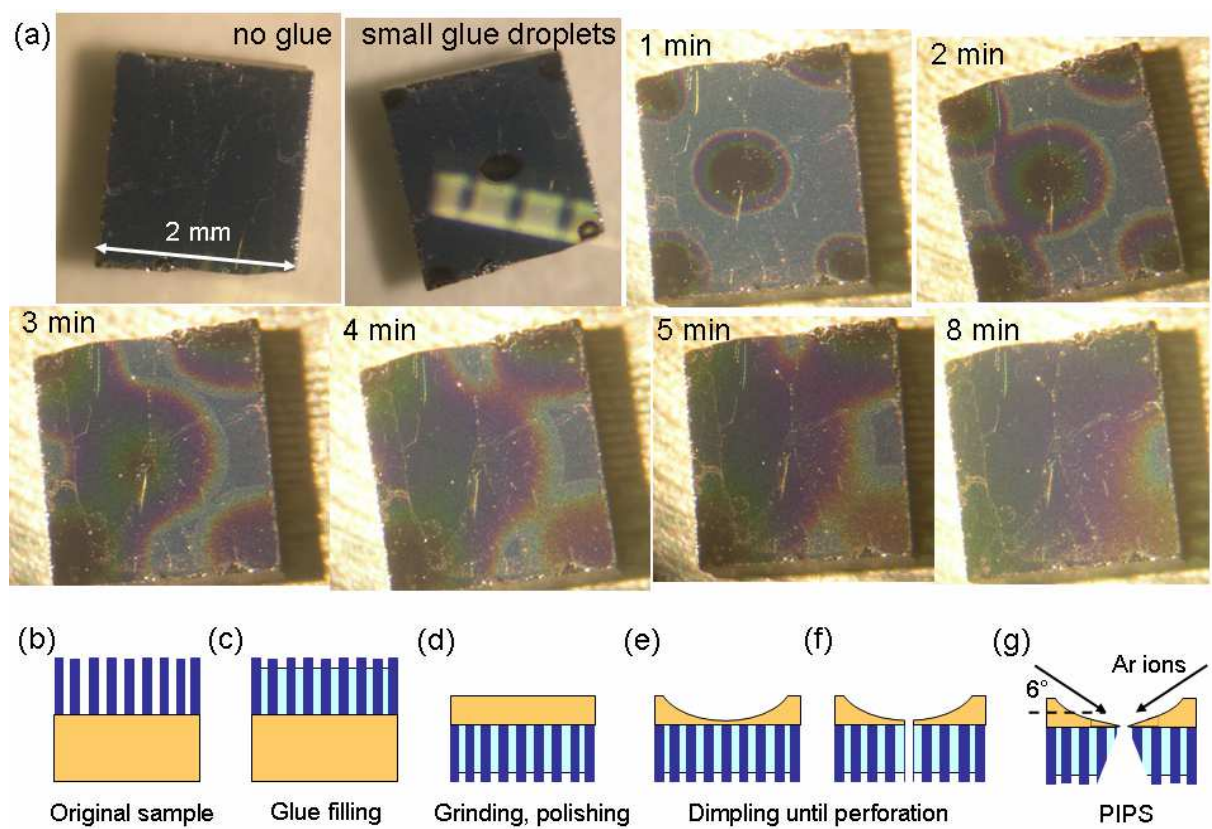

**Figure 2.** (a) Photographs illustrating the hand-coating procedure of the NWs with the cured epoxy GATAN G1: the deposited small glue droplets spread over the surface while heated on a hot-plate**.** (b) - (g) Schematic representation of the sample at different stages of the plan-view TEM preparation process (not to scale).

RESULTS AND DISCUSSION

**InN NWs case study: Native oxide investigation**

One of the most important characteristics of the InN NWs is the formation of a surface electron accumulation layer along the sidewalls (m-planes) (Calleja et al., 2007; Segura-Ruiz et al., 2009) and elucidating the origins of such electron accumulation would be of much help in order to create devices based on InN NWs. For that, we combined the information extracted from both cross-sectional and plan-view TEM. Our first finding was that, depending on the growth conditions, the InN NWs do not exhibit clearly defined lateral facets (cf. figure 1) resulting in rough sidewalls (Liliental-Weber et al., 2008). Rough sidewalls may also make the analysis by plan-view TEM complicated as the projected plan-view TEM micrographs of the NWs may not show well-defined crystallographic facets, but instead blurred sidewalls. We found that for a fixed V/III flux ratio, NWs grown at lower temperature presented rounded sidewalls and a higher degree of deformation together with a higher coalescence. When the NWs are grown under optimal temperature conditions (Grandal & Sánchez-García, 2005) the coalescence degree is reduced and well defined faceted NWs are obtained. In these samples, plan-view HRTEM micrographs [figure 3(b)] reveal a central InN core surrounded by an unintentionally formed thin (about 2-3 nm) and disordered shell layer. The interface, however, between this shell layer and the InN core is very sharp, showing that the facets of the InN core material are well-defined [figures 3(b) and 3(c)]. The unintentionally grown shell layer is composed of small nuclei. From the analysis of the Fourier power spectrum of the lattice images obtained in cross-section and plan-view geometry (figure 4), we extract that the pattern which is observed in both cross-section and plan-view geometries corresponds to the theoretical pattern of bcc $In_2O_3$, indicating that the shell layer is compatible with the formation of an unintentional $In_2O_3$ layer that is created as a consequence of the exposure to atmospheric air. Furthermore, the thickness of the native $In_2O_3$ shell varies slightly from sample to sample, observing that it is thicker in those samples that were longer exposed to air. This result is in agreement with the findings from (Ahn et al., 2010), although in their work they intentionally create an In$_2$O$_3$/InN core-shell structure using a rapid thermal oxidation process. In addition, saw tooth faceting has been reported in InN nanorods grown by hydride metal-organic vapor phase epitaxy and has been associated to the presence of In$_2$O$_3$ (Liliental-Weber et al., 2008). Also the formation of a surface In$_2$O$_3$ layer as a consequence of the natural aging process of InN quantum dots has already been reported (González et al., 2009).

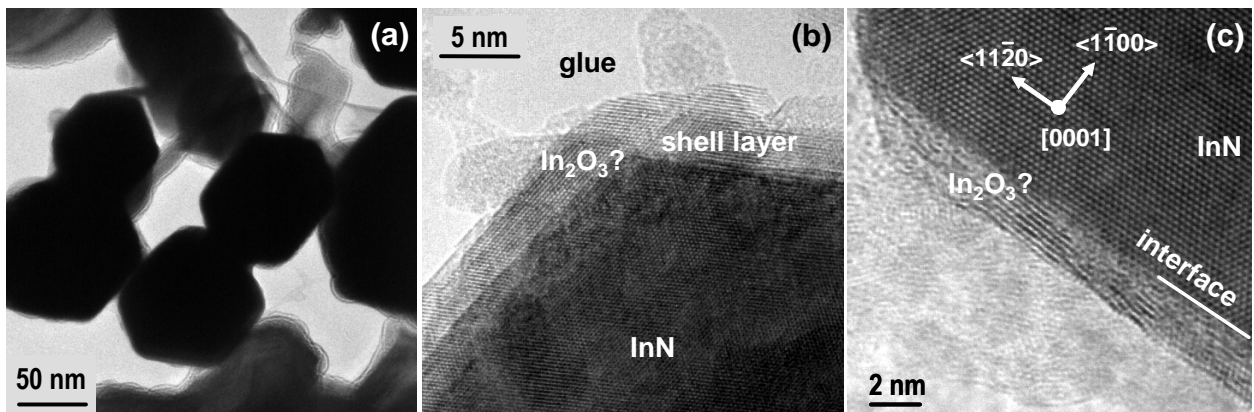

**Figure 3.** (a) Plan-view TEM micrograph of an overview of InN NWs prepared using the innovative procedure proposed in this work. (b)-(c) As observed in the HRTEM micrographs in plan-view geometry, a thin layer (~2 nm thick) forms around the InN NWs.

### Plan-view HRTEM

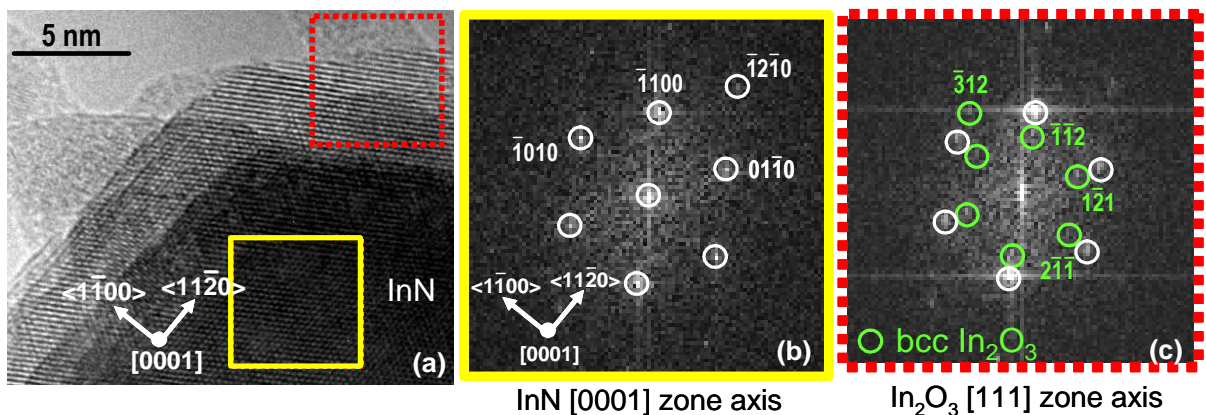

### Cross-sectional HRTEM

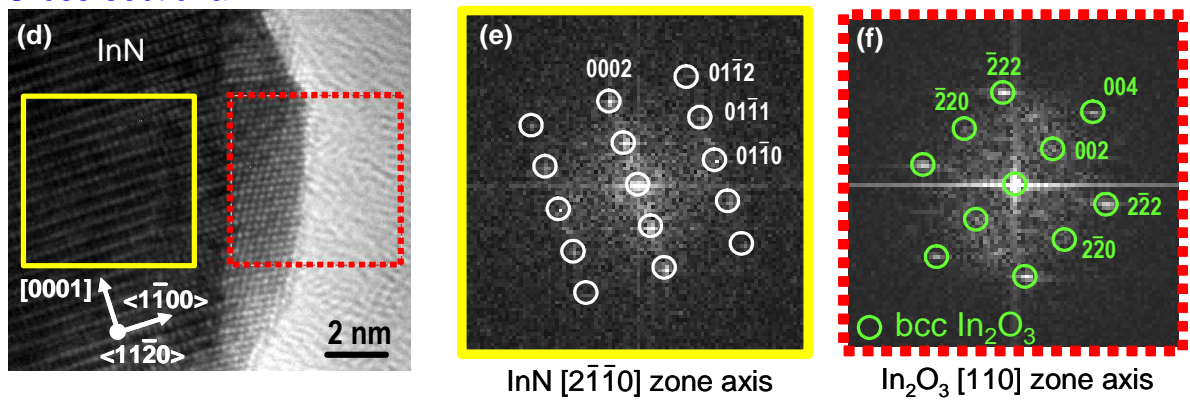

**Figure 4.** High-resolution TEM micrographs in (a) plan-view and (d) cross-section geometry of the InN NWs. Fourier power spectrum of the lattice images obtained in plan-view [(b)-(c)] and cross-section [(e)-(f)] geometry. The InN NWs and the In$_2$O$_3$ shell are epitaxially related as InN[0001]||In$_2$O$_3$[111] and InN <11-20> || In$_2$O$_3$ <110>.

Finally, note that electron accumulation also occurs at the surface of single-crystalline In$_2$O$_3$ (King et al., 2008, 2009), whose electronic structure and properties are similar to those of the other In-containing compounds InN (Mahboob et al., 2004) and InAs (Noguchi et al., 1991).

**InN NWs case study: Interface and Plastic relaxation**

In previous works, we have shown that when InN NWs are epitaxially grown on an AlN buffer layer, the strain between the AlN buffer and the InN NWs is relieved by an array of periodically spaced misfit dislocations (MDs) (Grandal et al., 2007). These dislocations are present at the InN/AlN interface, however, cross-sectional TEM observations do not provide a precise idea of the dislocation character and their actual two-dimensional arrangement. Taking advantage of our developed plan-view TEM technique for NWs, in particular to achieve very thin foils, we are able to observe a superposition of the HRTEM contrast of AlN with the aligned lattice of InN forming a Moiré-like pattern [figure 5(a)]. This Moiré-like structure with its six-fold symmetry reflects the network of misfit dislocations in symmetry and local atomic arrangement. Introducing an aperture to select one reflection, e.g. {1-100}, the corresponding Moiré fringes are formed in the image plane [see figure 5(b)]. The analysis of the Moiré fringes in plan-view TEM configuration can be used for a rough estimate of the degree of in-plane plastic relaxation of individual NWs. In particular, by measuring the distance between the fringes and using the expression for the spacing of translational Moiré fringes ($D_m$) given by:

$$D_m = \frac{d_{\{1-100\}}^{NW} d_{\{1-100\}}^{AlN}}{d_{\{1-100\}}^{NW} - d_{\{1-100\}}^{AlN}} \qquad (1)$$

where $d^{AlN}$ corresponds to the {1-100} plane spacing in the relaxed AlN buffer layer, we get $d^{NW}$ which is the {1-100} plane spacing in the InN NW (Kehagias et al., 2005; Lozano et al., 2006). From the plan-view HRTEM micrograph in figure 5(b), we measure an average Moiré fringes spacing of $D_m = 2.21$ nm. Using the literature value for $d^{AlN} = 0.2695$ nm, the distance of the {1-100} planes of InN NWs is determined to be $d^{NW} = 0.307$ nm. Based on that, we can estimate the amount of strain relieved by the network of MDs from the estimate of the plastic relaxation δ in the nanostructure (Kehagias et al., 2005; Lozano et al., 2006):

$$\delta = \frac{d^{AlN}_{\{1-100\}} - d^{NW}_{\{1-100\}}}{d^{AlN}_{\{1-100\}} - d^{InN}_{\{1-100\}}} \qquad (2)$$

where $d^{InN} = 0.3071$ nm corresponds to the {1-100} plane spacing in fully relaxed InN. Substituting values yields a δ ~ 99% plastic relaxation. Hence, InN NWs grown on an AlN buffer layer are almost completely relaxed by the formation of a MD network at the interface, further supporting the "strain-free" character of these NWs.

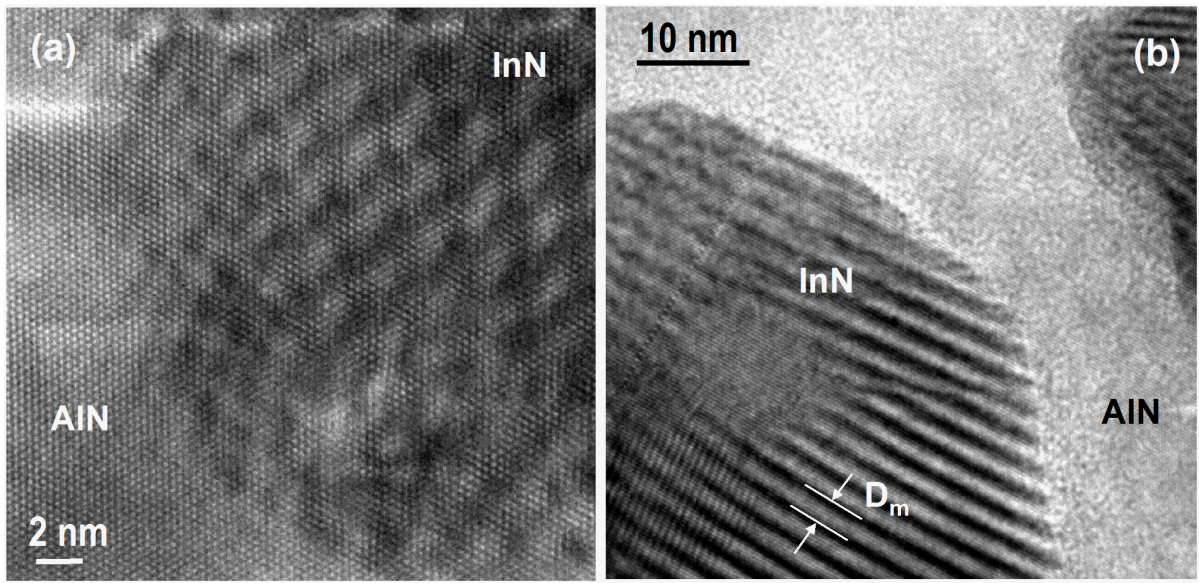

**Figure 5.** (a) Plan-view HRTEM image of an individual InN NW on an AlN buffer layer. Because of the very thin TEM foil, we are able to observe a superposition of the HRTEM contrast of AlN with the aligned lattice of InN forming a Moiré-like pattern with six-fold symmetry, which reflects the network of misfit dislocations at the InN/AlN interface. (b) Moiré fringes due to the lattice difference between an individual InN NW and the AlN buffer layer.

**Coalescence of InN nanowires**

Finally, we observed that in samples grown on bare Si(111) substrates, for the same V/III ratio, the NWs diameter increases as the temperature is reduced yielding to a higher degree of coalescence. In this situation, for NWs grown at low temperature, it is possible to observe the boundary between two coalesced NWs with atomic resolution [figure 6(a)]. As deduced from the analysis of the Fourier spectrum [inset in figure 6(a)], the two wires are tilted by about 15° in this case.

In the case of NWs grown on AlN-buffered Si(111) substrates at the optimum temperature, i.e. lower coalescence, it is also possible to study the properties of the underneath AlN layer, employed to improve the epitaxial alignment with the Si substrate and the nucleation of the InN NWs on top [figure 6(b)]. Plan-view TEM could also help to study the early stages of the NWs formation, including the stable island formation and the lateral (radial) growth (Consonni et al., 2011). Furthermore, plan-view TEM may provide crucial information concerning the mechanisms prevailing at the interface formation in NW core-shell structures, for instance. In particular, it is shown that a local two-dimensional island-mediated growth process dominates the interface evolution in planar heterostructures and determines its chemical interface width (Luna et al., 2012). Similar processes are also expected to occur in NWs, what would provide significant information about the dominant growth mechanism (Luna et al., 2012) as well as provide a precise control of the interfaces in heterostructures at NWs.

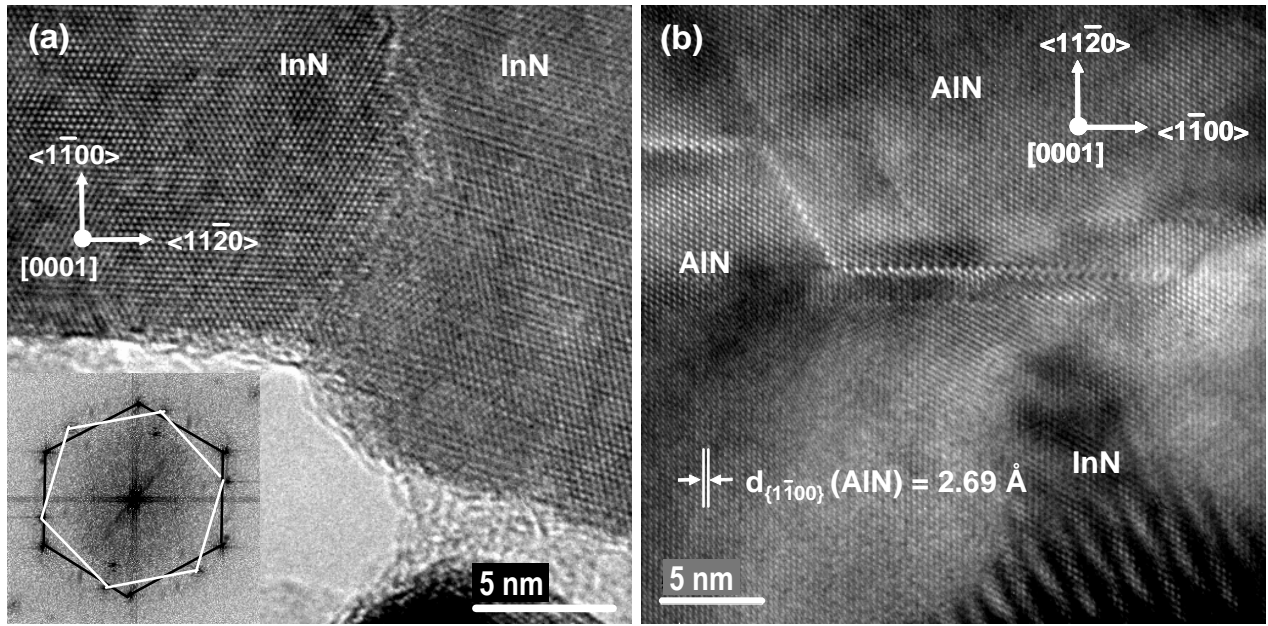

**Figure 6.** Plan-view HRTEM micrograph of (a) two coalesced InN NWs grown on bare Si and (b) of the AlN buffer layer employed to improve the epitaxial alignment of the InN NWs with the Si substrate.

CONCLUSIONS

In summary, we have explored different alternatives based on the conventional preparation tools and procedures for the plan-view TEM specimen preparation of NW semiconductor heterostructures. As far as HRTEM is concerned, we found that the best results are obtained using a refined method, based on the conventional procedure for thin-film plan-view TEM preparation but specifically modified for the NW morphology. The critical aspects are the prior mechanical stabilization of the NWs and the minimization of the ion-milling process after dimpling the samples until perforation. The combined analysis by plan-view and cross-sectional TEM of InN NWs allows the investigation of material aspects such as the formation of a native oxide layer. Further investigations may help in the controlled formation of the $In_2O_3$ layer and in determining its impact on the optical and electrical properties, i.e. the origin of the electron accumulation layer. In addition, for NWs grown on an AlN buffer, estimations of the plastic relaxation (99%) released by a MD network at the interface InN-AlN confirm the strain-free nature of such NWs. Finally, crucial information on the growth mechanisms and strain relaxation processes of core-shell NWs could be obtained from plan-view TEM observations.

ACKNOWLEDGEMENTS

The authors acknowledge A. Pfeiffer and D. Steffen for technical assistance with TEM, and A.-K. Bluhm for SEM observations. We acknowledge Uwe Jahn for a critical reading of the manuscript.